# Stability of Polarized States for Diamond Valleytronics

J. Hammersberg[1], S. Majdi[1], K.K. Kovi[1], N. Suntornwipat[1], M. Gabrysch[1], D.J. Twitchen[2] and J. Isberg[1*]

[1] *Division for Electricity, Uppsala University, Box 534, S-751 21 Uppsala, SWEDEN*

[2] *Element Six Ltd., Kings Ride Park, Ascot, Berkshire, SL5 8BP, UK*

**Abstract**

The stability of valley polarized electron states is crucial for the development of *valleytronics*. A long relaxation time of the valley polarization is required to enable operations to be performed on the polarized states. Here we investigate the stability of valley polarized states in diamond, expressed as relaxation time. We have found that the stability of the states can be extremely long when we consider the symmetry determined electron-phonon scattering. By Time-of-Flight measurements and Monte Carlo simulations, we determine electron-phonon coupling constants and use these data in order to map out the relaxation time temperature dependency. The relaxation time can be microseconds or longer below 100K and 100 V/cm for diamond due to the strong covalent bond, which is highly encouraging for valleytronic applications.

* Corresponding author: jan.isberg@angstrom.uu.se

*Intro*

In today's electronic circuits, information is encoded by the presence or absence of charge. Information processing therefore relies on the rapid redistribution of charge, which invariably requires energy, resulting in heat loss and thereby imparting a fundamental limit to the maximal switching rate. However, there are alternative concepts for processing information that rely on other properties of electrons than their charge. An example of a technology that does not exclusively rely on the charge is spintronics, where the spin angular momentum is utilized as a carrier of information. Another, relatively new concept, is *valleytronics* in which information is carried by the valley degree of freedom in a semiconductor where electrons occupy multiple degenerate conduction band minima (valleys)[1–4]. In addition, achieving successful manipulation of valley states could open up the possibility of valley-based *quantum* computing[5,6], either directly with valley-based qubits or alternatively via the spin-orbit interaction (SOI) using spin-based qubits.

Materials with multiple degenerate conduction band minima are, for example, graphene, $MoS_2$ and several bulk semiconductors such as silicon and diamond. In both silicon and diamond the conduction band has six degenerate valleys. In order to make useful devices, the electrons located in these valleys must retain their valley polarization for a long enough time to allow manipulations to be performed on them. Recently, it has been shown[7–9] that it is possible to create and detect electrons in a given valley in monolayer $MoS_2$ by pumping and detecting circularly polarized light. The valley polarization was in this case retained ~1 ns[8]. Our group has recently reported that the relaxation time of valley polarized states in diamond at liquid nitrogen temperature can be longer than 300 ns[10]. In this paper we investigate the stability of valley polarized states in more detail by considering the electron-phonon deformation potential coupling in Monte Carlo (MC) simulations. The deformation potential constants for diamond are determined from experiments by Time-of-Flight measurements. Knowing the strength of the deformation potential coupling, we can determine the relaxation time τ as a function of temperature and the applied electric field from MC simulations. Here we define τ from the expression $\frac{dn_\Delta}{dt} = -\frac{n_\Delta}{\tau}$ where $n_\Delta$ is the excess electron density in a given valley under non-equilibrium conditions.

Materials such as graphene, carbon nanotubes and diamond are all based on carbon and they have been the subject of intense research during the last decade. Diamond is well known for its exceptional hardness, but it is less well known that diamond also can conduct heat six times more efficiently than copper, that it can withstand very high electric fields without breakdown and that it can be doped to become electrically conducting. The exceptional properties of diamond together with the progress in thin-film synthesis, using chemical vapor deposition techniques[11,12], have created a strong interest in diamond as an electronic, photonic and spintronic device material. Possible applications range from nanoscale magnetic sensors[13] and single photon sources[14] to quantum computing.[15,16] It has been predicted[17] that diamond will be *the* material of choice for quantum applications; the reason being the very long spin relaxation time and the large band gap (5.5 eV) that allow control of optically active impurities. Now the additional possibility to exploit the valley degree of freedom in diamond for new types of qubits has evolved with the observation of relatively long-lived valley polarized electron states in diamond[10].

To study valley polarized currents we use the Time-of-Flight (ToF) measurement technique. In the ToF technique the motion of free carriers in an electric field induces a current that can be measured at contacts on the sample surface by an external circuit, see figure 1. The induced current is related to the electron motion through the Shockley-Ramo theorem[18,19]. This theorem states that the instantaneous current $I$ induced on a conductor A due to the motion of a charge carrier in a sample is $I = q\vec{v} \cdot \vec{E}$, where $q$ is the carrier charge, $\vec{v}$ is the instantaneous velocity of the point charge and $\vec{E}$ is the electric field that would exist at the position of the charge under the assumptions that the charge is removed, the conductor A is held at unit potential and all other conductors are grounded. Applied to a parallel contact geometry, this implies that the current $I$ measured at the contacts for a moving point charge is given by $I = qv_x/d$, where $v_x$ is the velocity component normal to the contacts and $d$ is the contact spacing. Therefore, a bunch of electrons traversing a sample at constant velocity will induce a constant current that rapidly drops to zero when the transit is complete. Similarly, several bunches of electrons moving at different speeds will induce a current that drops in well-defined steps, corresponding to the arrival of the bunches at the receiving contact. This can be used to detect electrons with different valley polarizations because of the difference in drift velocity anisotropy for electrons in different valleys.

In our experiment electron-hole pairs are generated by short (3ns FWHM) UV pulses from a quintupled Nd-YAG laser with 10 Hz repetition frequency and a wavelength of 213 nm. This wavelength corresponds to a photon energy slightly above the bandgap energy of diamond (5.47 eV). Several interference filters block lower harmonics and neutral density filters allow reducing the intensity to any desired level. The holes are directly collected by the cathode contact and the electrons are swept through the sample toward the anode by the applied electric field. To detect the induced current we use a low-noise broadband current amplifier, with a bandwidth of 1 GHz and a gain of 24 dB, together with a digital storage oscilloscope (DSO). The bias on the contacts is applied using a 50 μs pulser via a bias-tee. The short pulsed bias ensures capacitive voltage distribution across the sample and avoids undesirable sample charging. Two single-crystalline high-purity samples with a thickness of 490 and 510 μm, synthesized by Element Six Ltd. were chosen for this study. The material was deposited homoepitaxially on specially prepared high-pressure high-temperature (HPHT) synthetic diamond substrates in a microwave plasma-assisted chemical vapor deposition (CVD) reactor. After synthesis, the epitaxial overlayers were separated from their HPHT diamond substrate by laser cutting and polished to give freestanding plates (4.5×4.5 mm) with a nitrogen impurity concentration below $5\times10^{14}$ cm$^{-3}$. Semitransparent contacts were deposited on two opposite [100] faces of the samples. The semitransparent contacts make it possible to apply both a homogenous electric field and to create electron-hole pairs in the vicinity (to a depth of a few micrometers) of the illuminated surface of the sample, due to the strong absorption of the 213 nm light. The samples were mounted in a ceramic chip carrier, wire bonded and placed in a liquid helium cooled Janis ST-300MS vacuum cryostat with UV optical access. The temperature was monitored using a LakeShore 331 temperature controller with a calibrated TG-120-CU-HT-1.4H GaAlAs diode sensor in good thermal contact with the sample. Fig. 1 illustrates the schematics of the ToF setup used in this measurement.

Monte Carlo

The conduction band structure of diamond is similar to that of silicon, with six degenerate conduction band valleys oriented along the Δ-line, i.e. in (100)-directions, 76% of the way to the X-point of the Brillouin zone (BZ). Electrons in these valleys have a longitudinal effective mass[20] 1.56 $m_0$ and a transversal effective mass 0.28 $m_0$. Consequently, there is a strong anisotropy in the transport properties of valley-polarized electrons.

The phonon dispersion of diamond is also similar to that of silicon, since both are face center cubic materials. However, the phonon energies are considerably higher in diamond due to the rigid covalent carbon-carbon bond. In general, phonon energies are approximately three times higher in diamond than in silicon. In silicon and diamond, intervalley phonon *f*-scattering (scattering between valleys on orthogonal axes) requires interaction with longitudinal-acoustic (LA) or transverse-optical (TO) phonon modes close to the K point at the BZ boundary. Due to the rigid covalent bonds, the K-point LA or TO phonon energies are exceptionally high in diamond, about 130 meV[21]. The corresponding value for silicon is 43 meV.

For the Monte Carlo simulations we use a simplified conduction band structure, consisting of six parabolic but anisotropic valleys. This simplification is adequate at the moderate electric fields considered in this article and it reduces the computational effort considerably, compared to full-band simulations. Acoustic phonon intervalley scattering is treated through inelastic deformation-potential interaction. The deformation potential[22] $D_A$ is not known to high accuracy in diamond and we have therefore used it as a fitting parameter. Intervalley scattering is also treated through deformation-potential interactions[22], where the deformation potential for *f*-scattering $D_f$ is a second fitting parameter. The probability for direct *g*-scattering (scattering between valleys on the same axis) is substantially lower than for *f*-scattering, due to a higher barrier ~ 165meV. Therefore, repopulation between valleys on the same axis is more likely to occur via two subsequent *f*-scattering events than from one single *g*-scattering event, making *g*-scattering irrelevant here. Other material parameters used in the simulations are given in table I.

| Parameter | Diamond | Silicon |
| --- | --- | --- |
| transition phonon energy for *f*-scattering $\omega_f$ | 130 meV[21] | 43 meV |
| transition phonon energy for g-scattering $\omega_g$ | 165 meV[21] | 60 meV |
| longitudinal acoustic phonon velocity $v_l$ | 17.52 km/s | 8.43 km/s |
| transversal acoustic phonon velocity $v_t$ | 12.82 km/s | 5.84 km/s |
| acoustic phonon deformation potential $D_A$ | **(12.0 eV)** | 9.0 eV[23] |
| f-scattering deformation potential $D_f$ | **(4 ×10$^8$ eV/cm)** | 3 ×10$^8$ eV/cm[23] |
| g-scattering deformation potential $D_g$ | --- | 8 ×10$^8$ eV/cm[23] |
| longitudinal effective mass $m_l/m_0$ | 1.56[20] | 0.98 |
| transversal effective mass $m_t/m_0$ | 0.28[20] | 0.19 |

Table I. Parameter values used in the Monte Carlo simulation. Values are taken from [24] unless otherwise indicated. Values in boldface are results from the fitting procedure described in the text, rather than input parameters.

The acoustic phonon scattering rate can be written[21]:

$$P_{ac}(\vec{k},\vec{k}') = \frac{q\pi D_A^2}{V\rho v}\left\{(\exp(\frac{\hbar \kappa v_l}{kT})-1)^{-1} + \tfrac{1}{2} \pm \tfrac{1}{2}\right\} \cdot \delta(E_c(\vec{k}') - E_c(\vec{k}) \pm \hbar \kappa v)$$

where $\vec{k}$ ($\vec{k}'$) is the initial (final) hole wave vector, $\rho$ the density, and $\kappa$ length of the phonon wave vector. The upper sign is taken for phonon absorption and the lower sign for phonon emission. Similarly, the f-scattering rate is given by[21]:

$$P_f(\vec{k},\vec{k}') = \frac{4\pi D_f^2}{\rho V \omega_f}\left\{(\exp(\frac{\hbar \omega_f}{kT})-1)^{-1} + \tfrac{1}{2} \pm \tfrac{1}{2}\right\} \cdot \delta(E_c(\vec{k}') - E_c(\vec{k}) \pm \hbar \omega_f)$$

The valley polarization relaxation time τ then equals the inverse f-scattering rate, τ = 1/$P_f$. For the implementation of scattering and final state selection in the MC simulation we closely follow the treatment in[21].

The values of the deformation potentials $D_A$ and $D_f$ in diamond given in table 1 were obtained by varying these two parameters to determine the best fits of the calculated current with measured current traces from the ToF experiments. This was done for a set of measurements taken in the temperature interval 10-80 K for electric fields in the range 100-1000 V/cm. Examples of experimental current traces and currents calculated by MC simulations are plotted in figure 2. Clear steps can be observed in the curve traces; this is due to the different valley polarizations[10]. Varying the value of $D_g$ within reasonable limits in the MC simulations has negligible influence on the calculated currents, so this parameter cannot be determined in this way. This is due to the substantially higher phonon energy required for *g*-scattering compared to *f*-scattering.

The good agreement between experiment and simulations, using only two fitting parameters, is encouraging and we can now calculate relaxation times via the electron-phonon deformation potentials. Figure 3 shows the relaxation times at a fixed field and varying temperatures in diamond and, for comparison, in silicon.

At low fields below 100 V/cm and low temperature, below 100 K, the relaxation time, is very long, even microseconds. At higher fields, substantial electron heating reduces the relaxation time. Figure 4 shows the relaxation times for diamond at 77 K as a function of the applied field. At fields > 600 V/cm the relaxation time for the hot valleys has dramatically been reduced (to less than nanoseconds) due to the rapid heating of the electrons by the E-field and followed by phonon emission in an *f*-scattering event. This typical behavior can be utilized to write and transfer the electrons in the hot valleys into the cold valleys parallel to the field, as suggested in[10]. The relaxation time for acoustic deformation potential intravalley scattering (scattering within the same valley) time $\tau_{ac}$, is much shorter, due to the lack of a barrier. The electron mobility is thus dominated by intravalley phonon scattering, while *g*-scattering is irrelevant and only *f*-scattering causes de-polarization.

From the MC simulations we can calculate the relaxation times at various temperatures and the relaxation time for *f*- and *ac*-scattering are on entirely different time scales. By considering only electron-phonon interaction allowed by symmetry considerations we can say that the states in the hot and cold valleys have microsecond lifetimes at low fields <100 V/cm and temperatures below 100 K.

Performing the same calculation on silicon we obtain very different results. For silicon we require, e.g., 40 K and 10 V/cm to obtain the same relaxation time, one microsecond, as in diamond at 100 K and 100 V/cm. Consequently, valley polarization has not yet been observed in bulk Si, although it has been reported in a Si quantum well at LHe temperature[25].

In these Monte Carlo simulations we have only considered the electron phonon interaction. Neutral and ionized impurity scattering and electron-electron scattering can also cause de-polarization. Retaining a long polarization relaxation time thus requires samples with low impurity concentration and also necessitates low carrier concentrations. The influence of ionized impurity scattering is minimized in our experiments by selecting the purest CVD diamond samples available. The concentration of the dominant impurity, nitrogen, is below $10^{13}$ cm$^{-3}$ in all samples, as determined by electron paramagnetic resonance (EPR). From previous studies[26], it is known that the concentration of charged impurities is less than $10^{10}$ cm$^{-3}$. A negligible carrier-carrier scattering rate is assured in our experiments by controlling the amount of carrier generation and thereby limiting the free electron density to below $10^{10}$ cm$^{-3}$. One should also consider the band-to-band recombination time which for indirect band gap materials such as diamond and silicon can be several microseconds.

Conclusions

We have studied the stability of valley polarized states in diamond using Time-of-Flight measurements and Monte Carlo simulations. Here, we only study electron-phonon interactions that are allowed by symmetry selection rules. The deformation potentials $D_A$ and $D_f$ are extracted from the Time-of-Flight data by fitting the Monte Carlo simulations to the experimental results.

By varying the temperature and bias voltage in the Monte Carlo simulations we have found that the stability of the polarized states depends strongly on temperature and field due to the electron-phonon interaction. The valley polarization relaxation time varies over several orders of magnitude when the temperature is increased.

At low temperature, below 100 K, and at low fields, below 100 V/cm, the relaxation time can be very long, even microseconds. Thus, in order to use diamond in valleytronic applications one should aim at low temperature applications. Scattering due to ionized impurities and electron-electron scattering may also cause depolarization. High purity and low carrier concentrations are therefore important factors.
The obtained stability of the polarized states and the conditions for the stability we found are nevertheless highly encouraging for valleytronic applications.

This work was supported by the Swedish Research Council (Grant no. 621-2012-5819) and the STandUP for energy strategic research framework.

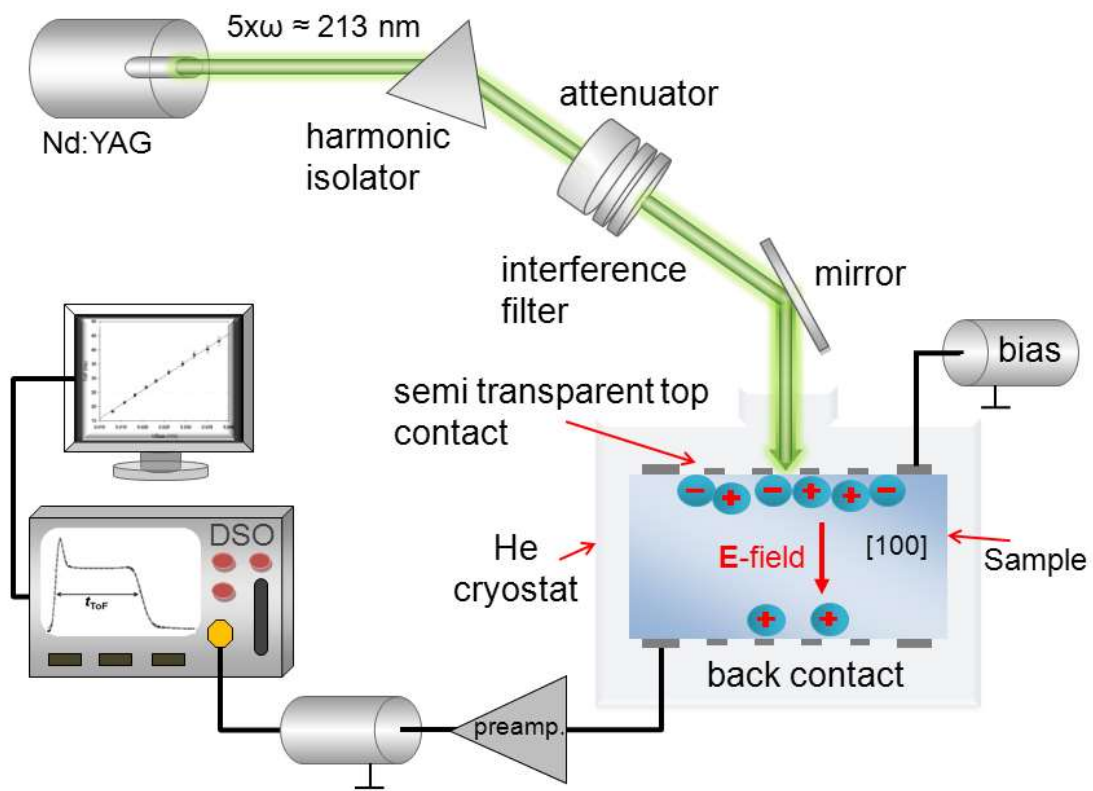

Figure 1. Schematics of the time-of-flight setup used in the experiment.

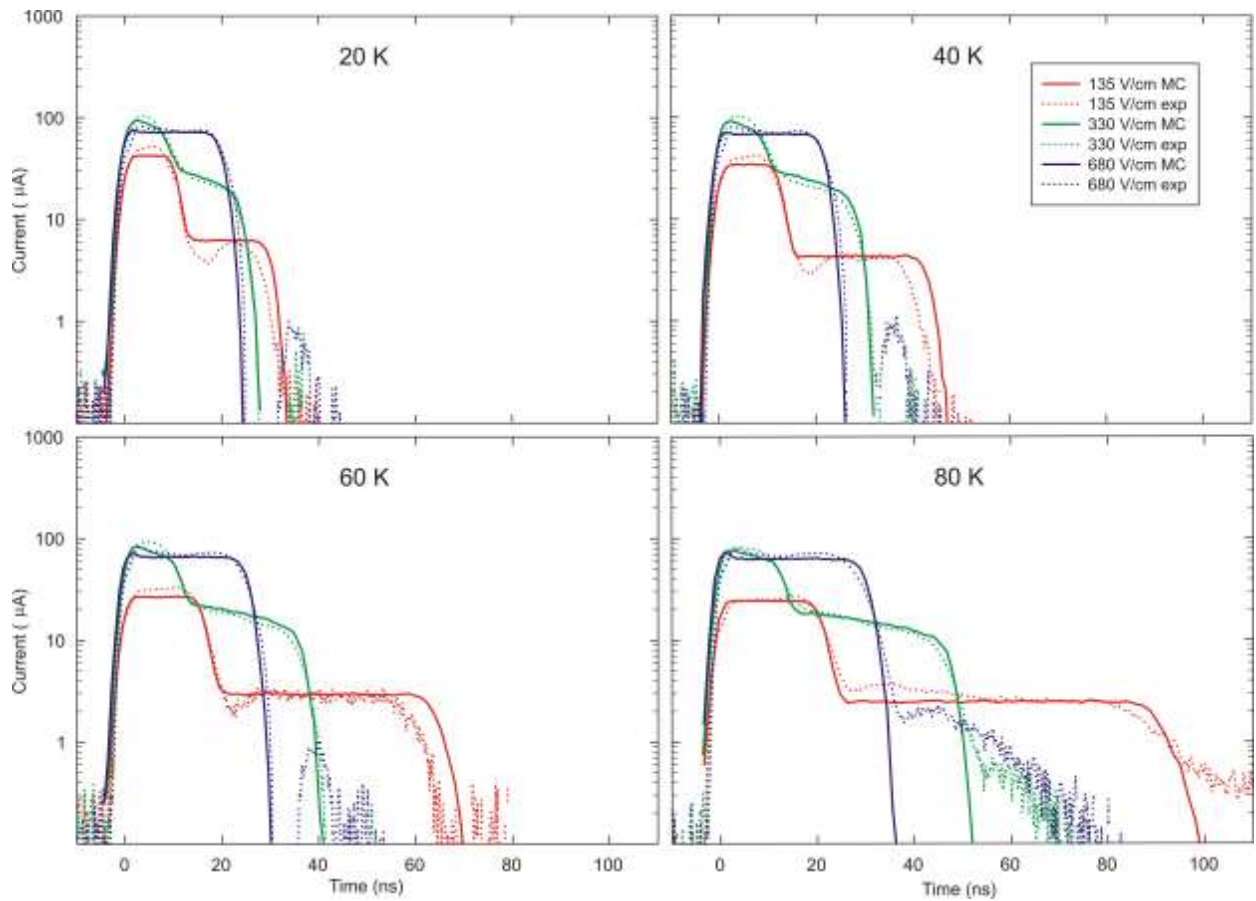

Figure 2. Examples of experimental time-of-flight data together with Monte Carlo simulations. In the Monte Carlo simulations the deformation potentials $D_A$ and $D_f$ are used as fitting parameters. The same $D_A$ and $D_f$ values are used for all temperatures and fields.

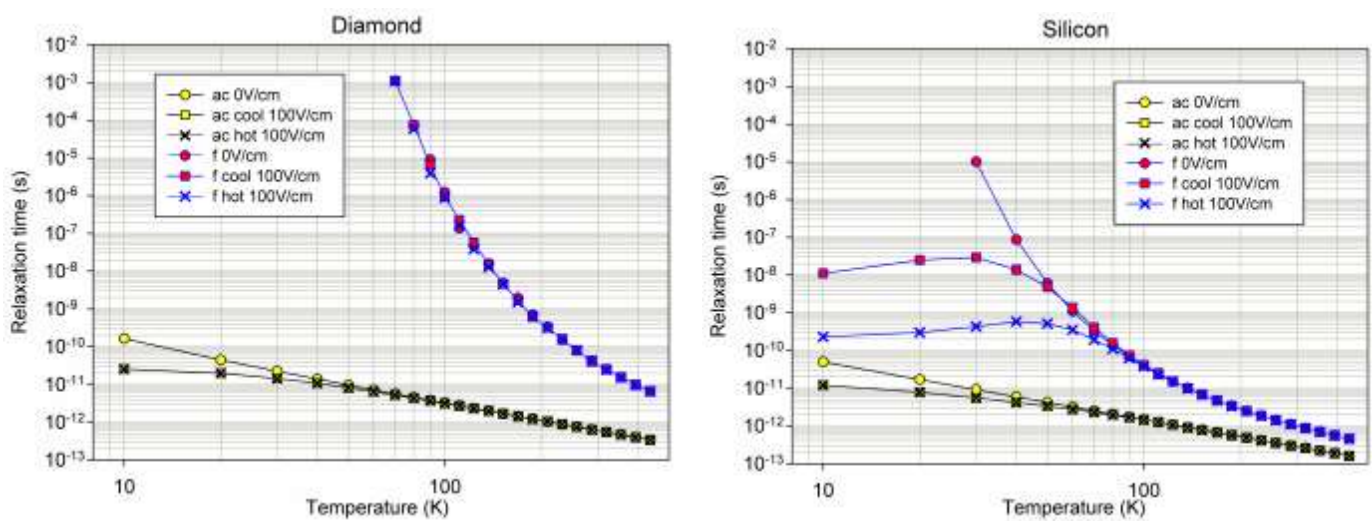

Figure 3. The relaxation time at 0 V/cm and 100 V/cm for *ac* (acoustic) intravalley scattering and *f* intervalley scattering for electron in cool and hot valleys. The stability of the valley polarized states in diamond is independent of the field up to 100 V/cm, whereas the polarized states in silicon are strongly affected by the field.

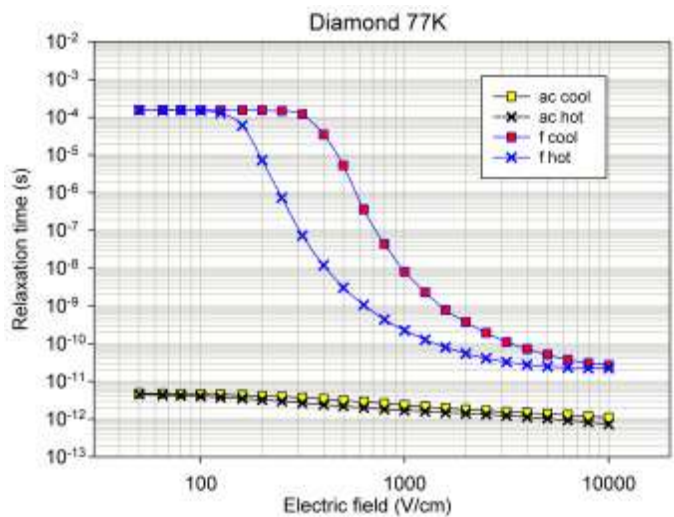

Figure 4. The relaxation time in diamond at 77 K for electric fields below 10000 V/cm for *ac* (acoustic) intravalley scattering and *f* intervalley scattering.